\documentstyle [12pt] {article}
\newcommand{\h}[1]{\mathop{\lambda}\limits_{#1}\ \!\!\!}

\newcommand{\tder}[2]{\frac{d{#1}}{d{#2}}}
\begin{document}
\begin{center}

\section*{NEW PATH EQUATIONS IN ABSOLUTE PARALLELISM GEOMETRY}
\end{center}
\begin{center}
{\bf   M. I. Wanas, M. Melek and M. E. Kahil}\\ {\it  Astronomy
and Meteorology Department, Faculty of Science, Cairo University,
                     Giza, Egypt. } \\

\end{center}
{\bf Abstract.} The Bazanski approach, for deriving the geodesic
equations in Riemannian geometry, is generalized in the absolute
parallelism geometry. As a consequence of this generalization
three path equations are obtained. A striking feature in the
derived equations is the appearance of a torsion term with a
numerical coefficients that jumps by a step of one half from
equation to another. This is tempting to speculate that the paths
in absolute parallelism geometry might admit a quantum feature.

\begin{center}
\section*{1. Introduction}
\end{center}

  It is well known that Riemannian geometry possesses two types of paths. The first
is the geodesic path which is given by the second order
differential equation, $$ \tder{U^{\mu}}{S} +
\{^{\mu}_{\alpha\beta}\} U^{\alpha} U^{\beta} = 0,  \eqno{(1)} $$
where $\{^{\mu}_{\alpha\beta}\}$ is Christoffel symbol, S is an
evolution parameter varies along the path and $U^{\alpha}$ is a
unit vector tangent to the path. The second is the null-geodesic
path given by the second order differential equation, $$
\tder{N^{\mu}}{\lambda} + \{^{\mu}_{\alpha\beta}\} N^{\alpha}
N^{\beta} = 0, \eqno{(2)} $$ where $N^{\alpha}$ is a null vector
tangent to the path, and $\lambda$ is the evolution parameter.

  In constructing his theory of general relativity (GR), Einstein has used Riemanian
geometry in which the first path taken to represent the trajectory
of a massive test particle, while the second taken to represent
the trajectory of massless particle moving in a gravitational
field.

  Bazanski (1977, 1989) has established a new approach to derive the equations
of geodesic and geodesic deviation simultaneously by carrying out
the variation on the following Lagrangian: $$ L_{B} = g_{\mu\nu}
U^{\mu} {\frac{D {\Psi}^{\nu}}{D S}}, \eqno{(3)} $$ where $U^{\mu}
= {\tder{x^{\mu}}{S}}$,  $g_{\mu\nu}$ is the metric tensor,
$\Psi^{\mu}$ is the deviation vector and $\frac{D}{D S}$ is the
covariant differential operator using Christoffel Symbol.

  In the last thirty years some problems appeared in GR, as a result of its
applications, especially in the domain of cosmology. Some authors
(cf. Mikhail and Wanas (1977), M\o ller (1978) and Hayashi and
Shirafuji (1978)) strongly believe that those defects may be
removed using a more general geometry than the Riemannian. A
possible candidate geometry for this purpose is the absolute
parallelism (AP) geometry (cf. Einstein (1929), Robertson (1932)
and McCrea and Mikhail (1956)).

  The aim of the present work is to find the possible paths in the AP-geometry
that can be considered as generalization of the paths in the
Riemannian geometry.

\section*{2. Path equations in AP-Geometry}

  In the AP-geometry, one can define four different affine connexions, Christoffel
symbols $\{^{\alpha}_{\mu\nu}\}$, the non-symmetric connexion
$\Gamma^{\alpha}_ {.\mu\nu}$ defined as a consequence of the
AP-condition, the dual connexion
$\tilde{{\Gamma}}^{\alpha}_{.\mu\nu} (=
{\Gamma}^{\alpha}_{.\nu\mu})$ and the symmetric part of the
non-symmetric connexion${\Gamma}^{\alpha}_{.(\mu\nu)}$. Using
these connexions one can define the following derivatives: $$
A^{\mu}_{~; \alpha} = A^{\mu}_{~, \alpha} +
\{^{\mu}_{\beta\alpha}\} A^{\beta}~~~, \eqno{(4)} $$ $$
A^{\stackrel{\mu}{+}}_{~~| \alpha} = A^{\mu}_{~, \alpha} +
{\Gamma}^{\mu}_{.\beta\alpha} A^{\beta}~~~,  \eqno{(5)} $$ $$
A^{\mu}_{~~| \alpha}  = A^{\mu}_{~, \alpha} +
{\Gamma}^{\mu}_{.(\beta\alpha)} A^{\beta}~~~,  \eqno{(6)} $$ $$
A^{\stackrel{\mu}{-}}_{~~| \alpha} = A^{\mu}_{~, \alpha} +
\tilde{{\Gamma}}^{\mu}_{.\beta\alpha} A^{\beta}~~~,  \eqno{(7)} $$
where $A^{\mu}$ is an arbitrary contravariant vector, and (,)
denotes ordinary partial differentiation. The coordinate
derivatives (4), (5), (6) and (7) are related to the parameter
derivatives by the following relations: $$ \frac{D A^{\mu}}{D S} =
A^{\mu}_{~;\alpha} U^{\alpha},  \eqno{(8)} $$ $$ \frac{{\bf{D}}
A^{\mu}}{{\bf{D}} S^{+}} = A^{\stackrel{\mu}{+}}_{~~| \alpha}
V^{\alpha}, \eqno{(9)} $$ $$ \frac{{\bf{D}} A^{\mu}}{{\bf{D}}
S^{0}} = A^{\mu}_{~~| \alpha} W^{\alpha},  \eqno{(10)} $$ $$
\frac{{\bf{D}} A^{\mu}}{{\bf{D}} S^{-}} =
A^{\stackrel{\mu}{-}}_{~~| \alpha} J^{\alpha},  \eqno{(11)} $$
where $S, S^{+}, S^{0}$ and $S^{-}$ are parameters varying along
the curves whose tangents are, respectively, $U^{\alpha},
V^{\alpha}, W^{\alpha}$ and$J^{\alpha}$ defined in the usual
manner.

  Now generalizing the Bazanski's Lagrangian (3) using (9), (10) and (11) we get
the following Lagragians: $$ L^{+} = \h{i}_{\alpha}~
\h{i}_{\beta}~ V^{\alpha}~ {\frac{{\bf{D}}{\xi}^{\beta}}{{ \bf{D}}
S^{+}}}~~~~~,  \eqno{(12)} $$ $$ L^{0} = \h{i}_{\alpha}~
\h{i}_{\beta}~ W^{\alpha}~ {\frac{{\bf{D}}{\zeta}^{\beta}}{{
\bf{D}} S^{0}}}~~~~~,  \eqno{(13)} $$ $$ L^{-} = \h{i}_{\alpha}~
\h{i}_{\beta}~ J^{\alpha}~ {\frac{{\bf{D}}{\eta}^{\beta}}{{
\bf{D}} S^{-}}}~~~~~,  \eqno{(14)} $$ where $\h{i}_{\mu}$ are the
tetrad vectors giving the structure of the AP-space;
${\xi}^{\beta}, {\zeta}^{\beta}$ and ${\eta}^{\beta}$ are the
vectors giving the deviation from the curves characterized by the
evolution parameters $S^{+}, S^{0}$ and $S^{-}$ respectively.

Carrying out the variation formalism on (12), (13) and (14),
noting that raising and lowering indices does not commute with the
differential operators used in (13), (14), we get respectively $$
{\frac{dV^\mu}{dS^+}} + \{^{\mu}_{\alpha\beta}\} V^\alpha V^\beta
= - \Lambda^{~ ~ ~ ~ \mu}_{(\alpha \beta) .} ~~~V^\alpha V^\beta,
\eqno{(15)} $$ $$ {\frac{dW^\mu}{dS^0}} + \{^{\mu}_{\alpha\beta}\}
W^\alpha W^\beta = - {\frac{1}{2}} \Lambda^{~ ~ ~ ~ \mu}_{(\alpha
\beta) .}~~~ W^\alpha W^\beta, \eqno{(16)} $$ $$
{\frac{dJ^\mu}{dS^-}} + \{^{\mu}_{\alpha\beta}\} J^\alpha J^\beta
= 0,  \eqno{(17)} $$ where $\Lambda^\alpha_{. \mu \nu}$ is the
torsion of space-time defined by $$ \Lambda^\alpha_{~ \mu \nu} =
\Gamma^\alpha_{. \mu \nu} - \Gamma^\alpha_{. \nu \mu}, \eqno{(18)}
$$

It can be shown that the 1st integrals of the equations (15), (16)
and (17) are given, respectively, by $$ g_{\alpha \beta} V^\alpha
V^\beta = V^2 ,  \eqno{(19)} $$ $$ g_{\alpha \beta} W^\alpha
W^\beta = W^2 ,  \eqno{(20)} $$ $$ g_{\alpha \beta} J^\alpha
J^\beta = J^2 ,  \eqno{(21)} $$ where V, W and J are constants
along the corresponding paths respectively. But since V, W and J
are scalars, one can conclude that these quantities are constants
in general (everywhere).

\begin{center}
\section*{3. Discussion}
\end{center}

  In generalizing the Bazanski's approach in the AP-geometry, three equations
of paths are obtained. These equations can be considered as
generalization of the geodesic equations in Riemannian geometry.
Moreover, equation (17) and its first integral (21) give rise to
the geodesic, (and null-geodesic upon reparameterization),
equation of Riemannian geometry. In this case the vector $J^\mu$
will be reduced to a unit vector (in the case of the geodesic) or
a null-vector (in the case of the null-geodesic). Thus, in
generalizing the Bazanski's approach, in the AP-geometry, we get
in addition to the geodesic and null- geodesic equations, two more
paths (15) and (16) contain torsion terms.

One can look at the three equations (15), (16) and (17) as
representing three path equations containing torsion terms with
different coefficients. The striking feature is that the
coefficients of the torsion terms are 1, ${\frac{1}{2}}$ and 0 in
the equations (15), (16) and (17), respectively. It is clear from
these equations that there is a jump equal to ${\frac{1}{2}}$,
from one path to another.

It is tempting to speculate that paths in the AP-geometry possess
some quantum features. The question, now, is: {\it What are the
physical trajectories, if any, that these paths represent ?} This
question might be answered in a forthcoming article.

\section*{References}
 Bazanski, S. L., (1977) Ann. Inst. H. Poincar\'e, A {\bf 27},
145.\\  Bazanski, S. L., (1989) J.Math. Phys., {\bf 30}, 1018.\\
Einstein, A., (1929), Sitz. Preuss Akad. Wiss. {\bf 1}, 1.\\
Hayashi, K., and Shirafuji,T., (1979) Phys. Rev. D{\bf 19},
3524.\\  McCrea, W. H. and Mikhail, F. I., (1956), Proc. Roy. Soc.
Lond. A{\bf 235}, 11.\\  Mikhail, F. I., and Wanas, M. I., (1977),
Proc. Roy. Soc. Lond. A{\bf 356}, 471.\\  M\o ller, C., (1978)
Mat. Fys. Medd. Dan. Vid. Selek., A{\bf 39}, 1.\\  Robertson, H.
P., (1932), Ann. Math. Princeton (2), {\bf 33}, 496.
\end{document}